\documentclass[
 reprint,
 superscriptaddress,
 amsmath,amssymb,
 aps,
]{revtex4-2}

\usepackage{graphicx}
\usepackage{bm}
\usepackage[version=4]{mhchem}
\usepackage{siunitx}
\usepackage{nicefrac}
\usepackage{xfrac}
\usepackage{braket}
\usepackage{verbatim}
\usepackage{etoolbox}
\usepackage[utf8]{inputenc}
\usepackage{mathtools}
\usepackage{soul,xcolor,colortbl}
\usepackage{hyperref} \hypersetup{colorlinks=true,citecolor=blue,linkcolor=blue,urlcolor=blue}
\usepackage{tabularx}
\usepackage{csquotes}
\usepackage{multirow}
\usepackage{longtable}
\usepackage{cellspace}
\usepackage{resizegather}
\usepackage{dcolumn}
\usepackage[normalem]{ulem}
\usepackage{physics}
\usepackage{placeins}
\usepackage{natbib}
\usepackage{orcidlink}

\begin{document}

\title{Towards Non-van der Waals 2D Topological Insulators}

\author{Mani Lokamani\,\orcidlink{0000-0001-8679-5905}}
\email[]{m.lokamani@hzdr.de}
\affiliation{Department of Information Services and Computing, Helmholtz-Zentrum Dresden-Rossendorf, 01328 Dresden, Germany}

\author{Gustav Bihlmayer\,\orcidlink{0000-0002-6615-1122}}
\affiliation{Peter Gr\"unberg Institut (PGI-1), Forschungszentrum J\"ulich and JARA, 52425 J\"ulich, Germany}

\author{Gregor Michalicek\,\orcidlink{0000-0003-4719-188X}}
\affiliation{Peter Gr\"unberg Institut (PGI-1), Forschungszentrum J\"ulich and JARA, 52425 J\"ulich, Germany}

\author{Daniel Wortmann\,\orcidlink{0000-0002-2248-1904}}
\affiliation{Peter Gr\"unberg Institut (PGI-1), Forschungszentrum J\"ulich and JARA, 52425 J\"ulich, Germany}

\author{Stefan Bl\"ugel\,\orcidlink{0000-0001-9987-4733}}
\affiliation{Peter Gr\"unberg Institut (PGI-1), Forschungszentrum J\"ulich and JARA, 52425 J\"ulich, Germany}
\affiliation{Institute for Theoretical Physics, RWTH Aachen University, 52062 Aachen, Germany}

\author{Rico Friedrich\,\orcidlink{0000-0002-4066-3840}}
\email[]{r.friedrich@hzdr.de}
\affiliation{Institute of Ion Beam Physics and Materials Research, Helmholtz-Zentrum Dresden-Rossendorf, 01328 Dresden, Germany}
\affiliation{Theoretical Chemistry, Technische Universit\"at Dresden, 01062 Dresden, Germany}

\date{\today}

\begin{abstract}
Non-van der Waals two-dimensional (2D) materials derived from strongly bonded non-layered crystals have recently emerged as a novel and rising platform for nanoscale research.
While uncovering and tuning their (opto-)electronic, catalytic, and magnetic properties has been the focus of intense research, the impact of spin-orbit coupling (SOC) onto their electronic structure has not yet been explored in detail.
Studying these effects is, however, particularly relevant due to their surface cation termination and the presence of heavy elements in several representative compounds.
Here, we investigate the effect of SOC onto the electronic structure of 2D AgBiO$_3$, NaBiO$_3$, and SbTlO$_3$.
While the first two systems show negligible band renormalization upon inclusion of relativistic effects around the band gap, SbTlO$_3$ showcases a large SOC induced splitting (\SI{229}{meV}) for the lowest conduction bands associated with a band inversion.
Substitution of Tl with Pb forming SbPbO$_3$ brings the band-inverted feature to the Fermi level.
Analysis of topological invariants and investigation of edge states of zig-zag and armchair ribbons within the \SI{200}{meV} gap confirms the topological nature of the band splitting.
Our work thus establishes a foundation for the systematic study of robust non-van der Waals 2D topological insulators.

\end{abstract}

\maketitle

\section{\label{sec:introduction}Introduction}

\noindent
2D topological materials with heavy-element-induced strong SOC hold great promise for next-generation electronics and quantum devices.
For a long time, research efforts on 2D systems were primarily focused on structures
separated~\cite{Novoselov_2005,Nicolosi_2013,Huang_2015_1,Velicky_2017}
as weakly bound 2D sub-units from anisotropic layered crystals held together by weak van der Waals (vdW) forces, such as graphene~\cite{Novoselov_2004} and
transition metal dichalcogenides~\cite{Manzeli_2017}.
The chemical exfoliation of the first atomically thin sheets from non-layered, \emph{i.e.}, non-vdW crystals established a new direction in the study of 2D systems.
Early examples include 2D WO$_3$ \cite{Guan_AdvMat_2017} as well as hematene and ilmenene extracted
from earth-abundant ores hematite $\alpha$-\ce{Fe2O3}~\cite{PuthirathBalan_2018_1}
and ilmenite \ce{FeTiO3}~\cite{PuthirathBalan_2018_2}.

In the years following the discovery of the first representatives,
numerous other systems were derived from non-vdW bonded compounds \cite{yadav_2018,Puthirath_2020,Kaur_ACSNano_2020,Puthirath_2021,Moinuddin_2021,Hu_2021,Yousaf_2021,Guo_2021,Gibaja_2021,Toksumakov_NPJ2DM_2022,Jiang_NatSyn_2023}
by a variety of experimental techniques~\cite{Kaur_AdvMat_2022,Balan_MatTod_2022}.
These investigations were also complemented by several data-driven and computational studies~\cite{ricofriedrich_2022_1,tombarnowsky_2023_1,Bagheri_JPCL_2023,Hu_NatComm_2023,Ono_PRB_2025,Zhong_npjcm_2025,Bagheri_CoM_2026,Barnowsky_ARXIV_2025}.
A great appeal of this new materials class is that,
unlike traditional 2D vdW-materials with saturated bonds,
non-vdW 2D compounds exhibit (re-)active surfaces terminated by cations
and are characterized by dangling bonds and surface states.
This unsaturated termination can be readily engineered to tune and enhance
their electronic as well as magnetic properties and potential functionality using adsorbates~\cite{Wei_2020,Barnowsky_2024}, deposition on substrates~\cite{Tantis_2023}, or by building non-vdW heterostructures~\cite{Nihei_ARXIV_2025}.

In addition to serving as a versatile platform for exploring
magnetism and spintronics in reduced
dimensions~\cite{Jin_2021,Mohapatra_2021,ricofriedrich_2022_1,Barnowsky_2024},
these systems have also revealed potential for advancements in (opto-)electronics
and catalysis~\cite{PuthirathBalan_2018_1,PuthirathBalan_2018_2,Toksumakov_NPJ2DM_2022}.
However,
the widespread adoption of 2D materials in general and non-vdW systems in particular for applications
is hindered because of (i) common internal defects
such as vacancies, impurities, and misaligned atoms,
as well as (ii) external defects that often stem from interactions with the environment~\cite{Giovannetti_2008,MELIOS2016273,Wood2014,Kumar_2022,C8TA10306B} -- leading to inconsistent adhesion and interference with adjacent materials~\cite{Lemme_2022}.

In this regard,
2D topological insulators (TIs)~\cite{Kou_2017,Zhang_2022,Weber_2024}
are appealing.
Their electronic structure
is characterized by band-splitting and band-inversion at high-symmetry $\mathbf{k}$-points~\cite{Zhu_2012,Vidal_2012},
and the magnitude of the splitting is largely
determined by the strength of the SOC of the constituent heavy elements.
Their surface states are preserved by time-reversal symmetry and are thus robust against disruptions~\cite{Analytis_2010,Zhang_2012},
including scattering from non-magnetic impurities and surface distortions.
Since these edge states are stable and topologically protected,
2D TIs have been proposed as candidates
for quantum circuits~\cite{Zhang_2023} and dissipationless electronics~\cite{Qing_2024}.

\begin{figure*}[ht!]
    \begin{center}
        \includegraphics[width=.99\textwidth]{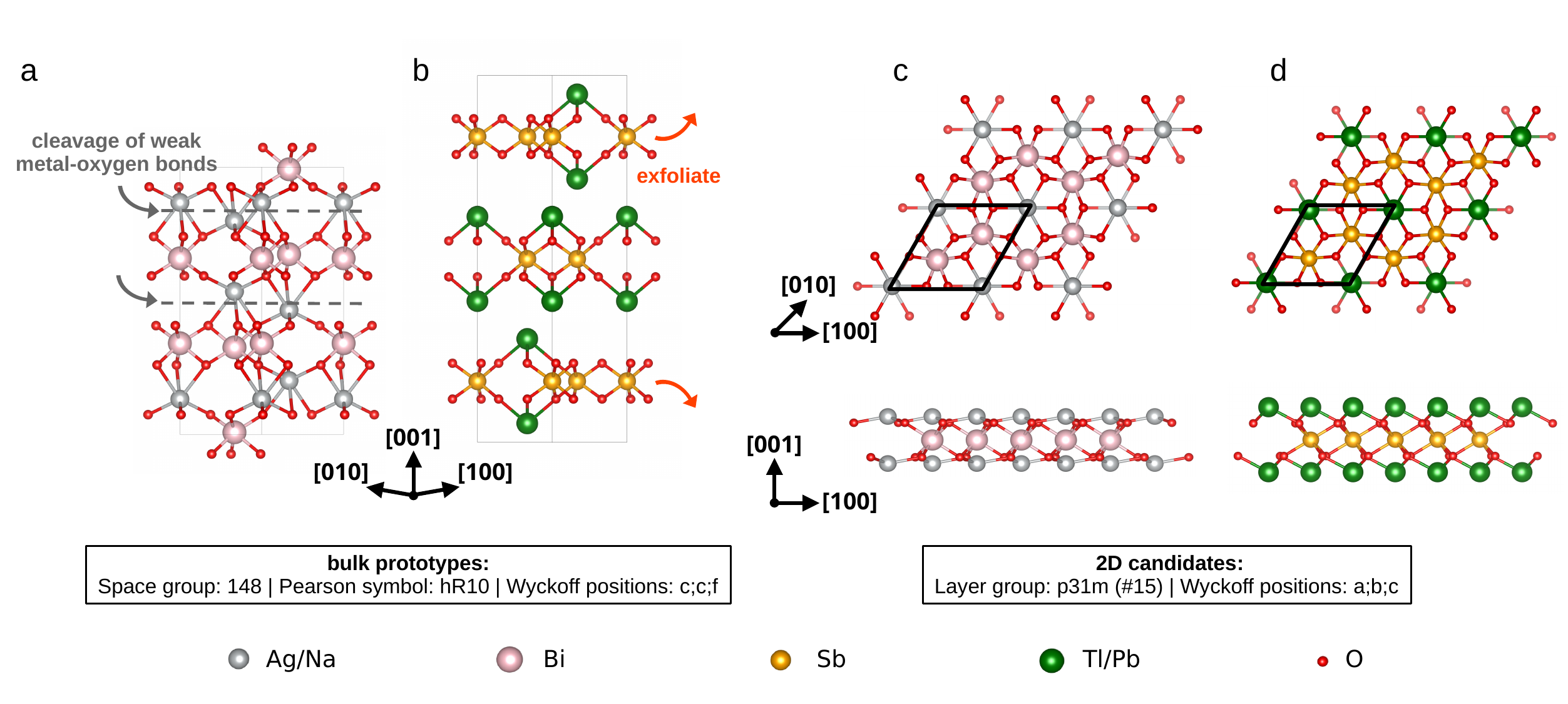}
    \end{center}
    \caption[schematics]{
    \textbf{Presentation of studied non-vdW 2D materials with respect to their bulk counterparts.}
    (a) Bulk system with short/long (strong/weak) metal-oxygen bonds.
    The long (weak) bonds define natural exfoliation planes \cite{ricofriedrich_2022_1}
    indicated using grey dashed lines perpendicular to the $\mathrm{[001]}$ direction.
    (b) Bulk system with exfoliable facets.
    Space group (SG), Pearson symbol (PS) and Wyckoff positions (WPs) of the structural prototype of both bulk systems
    are provided in the box.
    The compass indicating crystal directions applies to both subfigures.
    (c/d) Structure of non-vdW 2D materials \ce{AgBiO3} and \ce{NaBiO3}/\ce{SbTlO3} and \ce{SbPbO3} in top and side views.
    The in-plane unit cell of the 2D structures is indicated by the black frames.
    Layer group and Wyckoff positions of the structural prototype of the 2D candidates
    are indicated in the box.
    }
    \label{fig:schematics}
\end{figure*}

In this work, we focus on the previously outlined set of non-vdW 2D materials ~\cite{ricofriedrich_2022_1,tombarnowsky_2023_1} including systems with heavy elements such as Bi and Tl.
We specifically investigate the influence of SOC onto the electronic structure of 2D \ce{AgBiO3}, \ce{NaBiO3} and \ce{SbTlO3}.
While for the first two candidates, relativistic effects lead only to an insignificant modification of the band structure, for \ce{SbTlO3}, a large SOC-induced band splitting of \SI{229}{meV} is found in the conduction band manifold.
Element substitution of Pb for Tl allows for proper electron doping in \ce{SbPbO3} such that the band splitting is moved to the Fermi level.
Analysis of the orbital-projected band structures outlines that the splitting is associated with a prototypical inversion of the band character at the high-symmetry $K$- and $M$-points.
The calculation of the topological invariants and protected 1D conduction edge channels for both zig-zag and armchair ribbons finally confirms the topologically non-trivial nature of the surface states within the SOC-induced gaps.
The inclusion of relativistic interactions thus renders 2D \ce{SbPbO3} a robust TI.

\section{\label{sec:workflows}Methods}

\noindent
This work utilizes density functional theory (DFT) within the AiiDA-FLEUR~\cite{fleurWeb,Wortmann_fleur_zenodo,Huber_aiida_scid_2020,Uhrin_aiida_cms_2021} framework for accurate relativistic band structure calculations on top of
existing AFLOW~\cite{Oses_CMS_2023,Esters_CMS_2023,Divilov_HighEntAlloyMat_2025} workflows for standardized high-throughput \emph{ab initio} geometry optimization.
The relaxed input structures of \ce{AgBiO3}, \ce{NaBiO3}, and \ce{SbTlO3} are extracted from the dataset of non-vdW 2D materials created
in earlier studies~\cite{ricofriedrich_2022_1,tombarnowsky_2023_1}.
The initial structure of \ce{SbPbO3} is derived from \ce{SbTlO3} by replacing Tl by Pb
and subsequently relaxing both the ionic positions and the cell shape
with AFLOW~\cite{Oses_CMS_2023,Esters_CMS_2023,Divilov_HighEntAlloyMat_2025} and the Vienna \emph{ab initio} simulation package (VASP)~\cite{Kresse_1993_1,Kresse_1996_1,Kresse_1996_2}
with settings according to the AFLOW standard~\cite{Calderon_2015_1}.
The internal VASP precision is set to ACCURATE.
The automatic \textit{k}-point determination feature in AFLOW
is set to utilize \num{1000} \textit{k}-points per reciprocal atom
resulting in a $\Gamma$-centered $10 \times 10 \times 1$ grid for the 2D facet calculations.
Structural information regarding the layer group and Wyckoff positions of the 2D candidates are included in the Appendix.

The extracted Wyckoff positions are then utilized
to generate thin film geometries for FLEUR.
The self-consistent calculations in film mode~\cite{Krakauer_1979_1}
are combined with the downstream bandstructure and density of states (DOS) analysis
using the work-chains defined in AiiDA-FLEUR.
For these,
the $k$-point grid,
the {\small \textit{U}}-parameter for silver 4\textit{d}-states ($U = \SI{5.8}{eV}$, $J = \SI{0.0}{eV}$),
and the $k$-path for the band-structure calculations are inherited from the AFLOW standard workflow.
The kinetic energy cutoff $K_\mathrm{max}$ was set to
$4.5 \times \nicefrac{1}{a_0}$, where $a_0$ denotes
the Bohr radius.
The remaining simulation parameters were chosen by employing the profile \textit{moderate}
of the FLEUR input generator.

\begin{figure*}[ht!]
    \begin{center}
    \includegraphics[width=\textwidth]{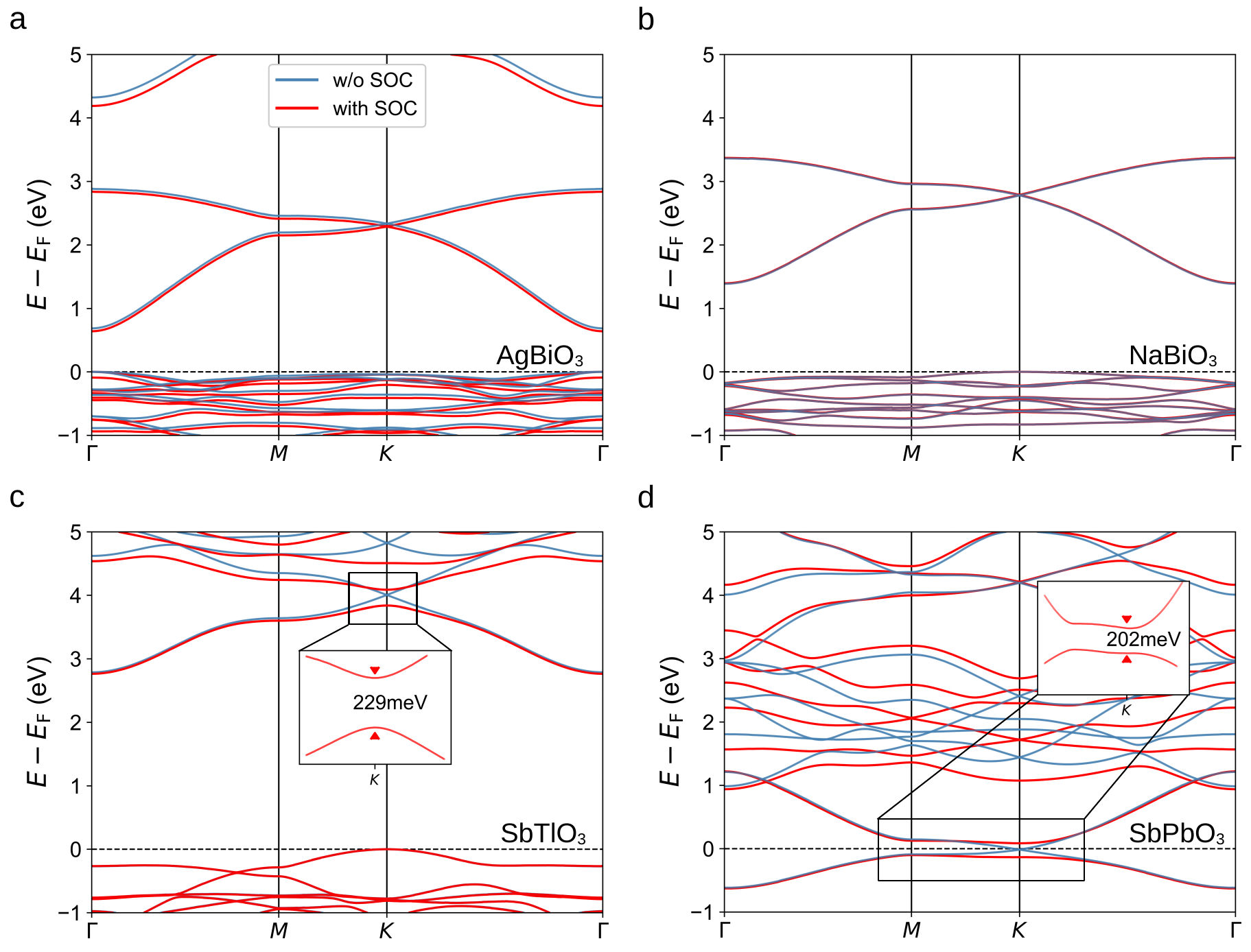}
    \end{center}
    \caption[bandstructureComparison]{
    \textbf{Bandstructures with and without SOC.}
    Bandstructures of \ce{AgBiO3} (a), \ce{NaBiO3} (b),
    \ce{SbTlO3} (c), and \ce{SbPbO3} (d) computed with and without SOC.
    The spin-orbit induced splitting for \ce{SbTlO3} and \ce{SbPbO3} at the high-symmetry $K$-point is enlarged in the respective sub-figure insets with the magnitude indicated in each case.
    }
    \label{fig:bandstructureComparison}
\end{figure*}

For the investigations of the topological properties of the Sb-based materials, we employed the wannier90 code~\cite{Mostofi_2014} to create maximally localized Wannier functions representing the $s$- and $p$-states of the Sb, Tl, and Pb and the $p$-states of O. We further used WannierTools~\cite{WT_2018} to calculate the edge states for semi-infinite ribbons.

\section{\label{sec:candidates}Results and Discussion}

\subsection*{Presentation of 2D candidates}

\noindent
The non-van der Waals 2D materials studied in this work can be derived
from bulk parent systems as shown in Fig.~\ref{fig:schematics}(a,b).
The first category of bulk systems giving rise to AgBiO$_3$ and NaBiO$_3$ sheets is characterized by anisotropic metal-oxygen bond strengths as outlined in Ref.~\cite{ricofriedrich_2022_1}.
These define natural exfoliation planes
perpendicular to the $\mathrm{[001]}$ direction to derive
2D sheets from their corresponding bulk structures as depicted in Fig.~\ref{fig:schematics}(a) using dashed lines.

The other flavor of bulk systems is comprised of exfoliable facets (see Fig.~\ref{fig:schematics}(b)),
that can be delaminated into individual sheets as in case of SbTlO$_3$.
While this setup is somewhat reminiscent of traditional 2D materials such as MoS$_2$ obtained from layered bulk crystals, we note that for our systems, the resulting 2D sheets are cation terminated which has been pointed out to be an important distinction of these non-vdW 2D systems~\citep{tombarnowsky_2023_1}.
The four derived non-vdW 2D candidates \ce{AgBiO3}, \ce{NaBiO3}, \ce{SbTlO3}, and \ce{SbPbO3}
studied in this work are shown in Fig.~\ref{fig:schematics}(c,d) in both top and side view.

\begin{figure*}[ht!]
    \begin{center}
    \includegraphics[width=\textwidth]{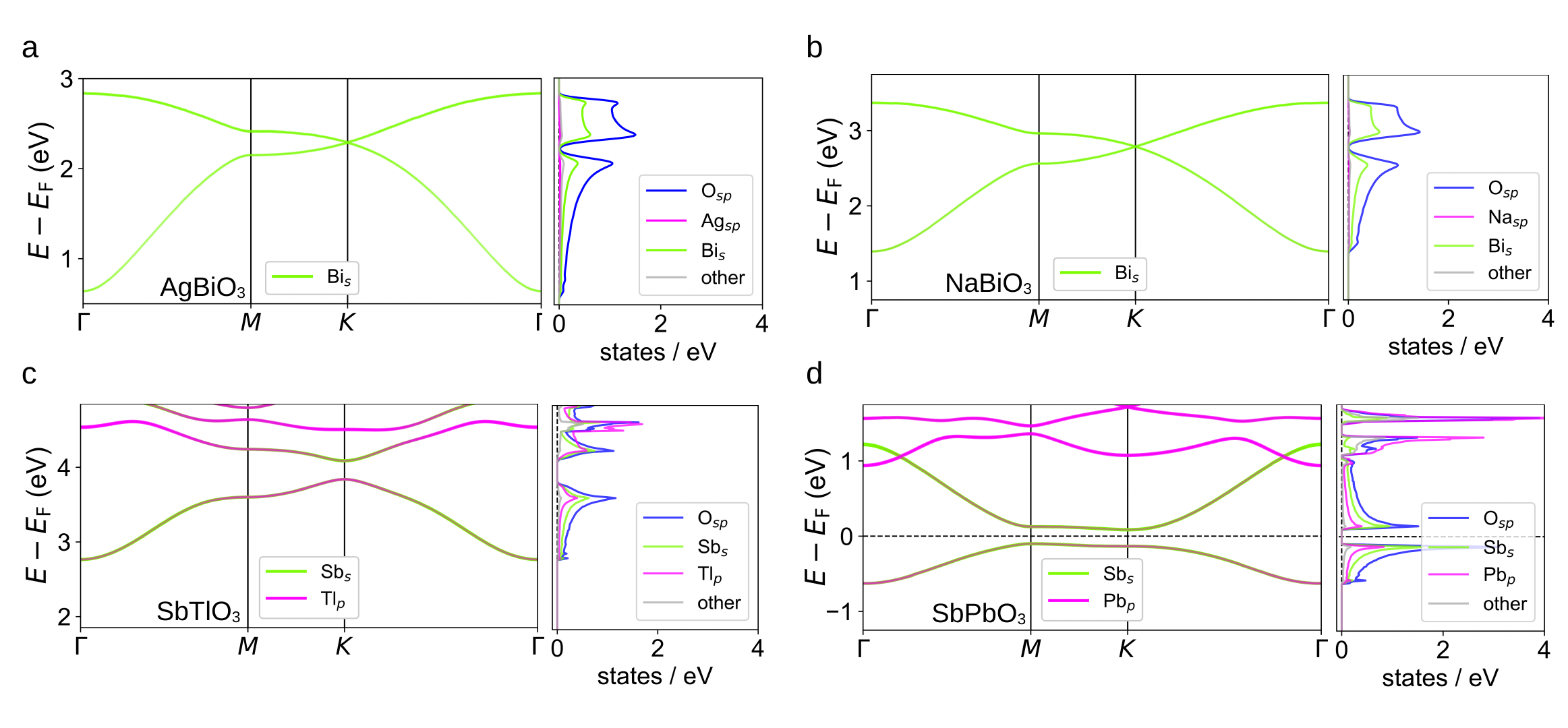}
    \end{center}
    \caption[orbitalProjectedBandstructure]
        {\textbf{Orbital-projected band structures.}
        Orbital-projected band structures
        and DOSs of
        \ce{AgBiO3} (a), \ce{NaBiO3} (b), \ce{SbTlO3} (c), and \ce{SbPbO3} (d)
        including SOC.
        In the band structures only the contributions from \textit{s}- and/or \textit{p}-states of the cations are highlighted.
        In addition, the DOS also shows contributions of \textit{s}- and \textit{p}-states from the oxygen atoms and
        the combined contributions from the remaining (other) states.
        In case of \ce{SbTlO3}, a band inversion is noticeable at the high-symmetry $K$-point,
        whereas in the case of \ce{SbPbO3}, band inversion is visible between the high-symmetry $M$- and $K$-points.
        }
    \label{fig:orbitalProjectedBandstructure}
\end{figure*}

\subsection*{\label{sec:accurateelectronicstructures}Band structure analysis}

\noindent
The 2D materials under consideration contain heavy elements such as bismuth,
thallium, and lead making them prototypical candidates for exploring the
impact of SOC on their electronic properties.
An overview comparison of the band structures computed with and without SOC
is presented in Fig.~\ref{fig:bandstructureComparison}.
For \ce{AgBiO3}, the inclusion of relativistic interactions results
in only a slight band renormalization
of the unoccupied states as well as the occupied states
close to the valence band maximum (see Fig.~\ref{fig:bandstructureComparison}(a)).
Similarly, for \ce{NaBiO3},
the inclusion of SOC has a negligible effect on the electronic structure,
as noticeable by the overlapping dispersions (Fig.~\ref{fig:bandstructureComparison}(b)).
However, for \ce{SbTlO3},
the inclusion of relativistic effects
leads to a gap opening at the high-symmetry $K$ point,
specifically for the conduction bands between \SI{2.7}{eV} and \SI{4.7}{eV}.
This band splitting at $K$ is substantial,
with a value of approximately \SI{229}{meV}
(see inset in Fig.~\ref{fig:bandstructureComparison}(c)).

To investigate whether this feature within the unoccupied manifold can be brought to the Fermi level,
the consequences of electron doping were investigated using element substitution.
As Pb provides one more valence electron than Tl and there are two cations of each type per unit cell, \ce{SbPbO3} becomes an appealing candidate.
The computed band structures in Fig.~\ref{fig:bandstructureComparison}(d) indeed confirm that a spin-orbit induced splitting of similar size of \SI{202}{meV} is now present at $K$ opening a sizeable gap right around the Fermi level.

To understand why the effect of SOC varies so significantly among the 2D candidates,
Fig.~\ref{fig:orbitalProjectedBandstructure} visualizes the orbital-projected band structure and DOS decomposed
into contributions from \textit{s}- or \textit{p}-states of the cations.
Additional contributions are included in the DOS,
specifically \textit{s}- and \textit{p}-states from the anions (oxygen),
as well as combined contributions from the remaining (other) states.
Notably,
the bands of interest for \ce{AgBiO3} and \ce{NaBiO3}
are composed of unoccupied \ce{Bi} \textit{s}-states
that do not contribute to SOC-effects due to zero angular momentum
as shown in Fig.~\ref{fig:orbitalProjectedBandstructure}(a,b).
This also fits to the Bi ${+5}$ oxidation state in these compounds leaving the Bi \textit{s}-orbitals as the first unoccupied levels.

However, in the case of \ce{SbTlO3} and \ce{SbPbO3} as
shown in Fig.~\ref{fig:orbitalProjectedBandstructure}(c,d)),
the considered bands are not only derived from \ce{Sb} \textit{s}-states,
but also include \textit{p}-character of \ce{Tl} and \ce{Pb},
which does contribute to SOC-effects.
For \ce{SbTlO3}, a prototypical band inversion from \ce{Sb} \textit{s}- to \ce{Tl} \textit{p}-character for the lower band and \emph{vice versa} for the upper one is clearly visible
at $K$.
For \ce{SbPbO3},
the band inversion is somewhat more spread between $M$ and $K$, still exhibiting a clear modulation of the band characters now right around the Fermi level due to the increased number of valence electrons.
These findings hint at a topological nature of the SOC-induced band splitting which we investigate in detail in the following.

\begin{figure*}[ht!]
    \begin{center}
    \includegraphics[width=\textwidth]{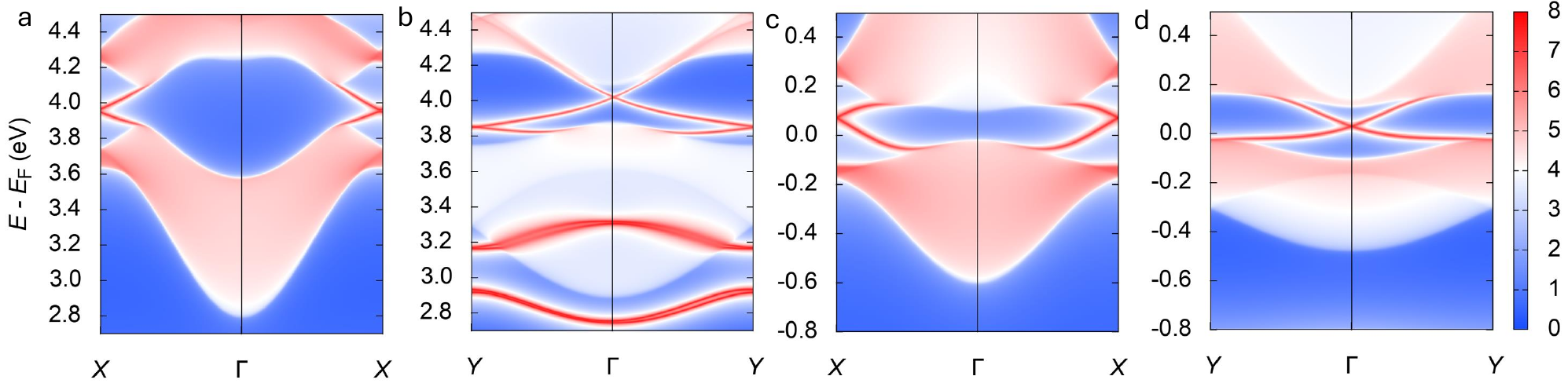}
    \end{center}
    \caption[topology]
        {\textbf{Edge states.}
        Shown are the edge-projected spectral functions of ZZ (a,c) and AC (b,d) ribbons of \ce{SbTlO3} (a,b) and \ce{SbPbO3} (c,d) as obtained from Wannier-function calculations. Red color indicates high density of states (shown in logarithmic scale) at the edges, visualizing the topological edge states of \ce{SbTlO3} between $3.9$ and $4.1$~eV, and for \ce{SbPbO3} between $-0.1$ and $0.1$~eV. Additional (trivial) edge states are visible in panel (b) around $3$~eV, for the corresponding ZZ edge (panel a), they are located around $2$~eV (not shown).
        }
    \label{fig:topology}
\end{figure*}

\subsection*{\label{sec:topologicalcharacter}Topological character and edge states}

We analysed the topological character of the Sb containing 2D materials by calculating the parity of the occupied states at the time-reversal invariant midpoints (TRIMs) and found that \ce{SbTlO3} is trivial ($\mathbb{Z}_2 = 0$) but adding one electron per formula unit moves the Fermi level into a gap where $\mathbb{Z}_2 = 1$ is found. This finding already suggests that \ce{SbPbO3} might be a 2D TI and DFT+SOC calculations confirm that $\mathbb{Z}_2 = 1$ in this case. By wannierization we reconfirmed this result and  calculated the edge states for semi-infinite ribbons. Two different edges can be considered, here the terms zig-zag (ZZ) and armchair (AC) edges are defined with respect to the Sb sublattice. The ZZ edge runs in $[100]$ direction (cf.\ Fig.~\ref{fig:schematics}), while the AC edge runs in $[\overline{1}20]$ direction.

As can be seen in Fig.~\ref{fig:topology} (a) and (b), edge states appear in the gap around $4$~eV for \ce{SbTlO3} that connect the upper and lower band edges. While the Dirac point is at the boundary of the 1D Brillouin-zone (BZ) for the ZZ edge (a), it appears also at the zone center for the AC configuration (b). In addition to these topology-induced edge states, some "trivial" states originating from broken bonds can be seen in panel (b). Also for the ZZ edge a flat band appears at approximately $2$~eV, however, these states might disappear when relaxations or reconstructions of the edges are taken into account~\cite{Kim_2018}.

Fig,~\ref{fig:topology} (c) and (d) show the topologically protected edge states of \ce{SbPbO3} ribbons with ZZ and AC edges, respectively. Also here the Dirac points are located at the BZ boundary in the former and BZ center in the latter case. These states are positioned in a rather large gap of $0.2$~eV that renders this 2D material a robust 2D TI.

\section{Conclusions}

\noindent
In this \emph{ab initio} study, we investigated the influence of SOC on the electronic structure of non-vdW 2D materials including heavy elements.
For both \ce{AgBiO3} and \ce{NaBiO3}, the influence of relativistic effects on the band structure around the gap is negligible due to the apparent $s$-character of the relevant Bi states.
In contrast, for \ce{SbTlO3}, SOC induces a substantial band-splitting of $\sim 229$~meV
at the high-symmetry \textit{K}-point within the conduction band manifold.
This splitting is accompanied by a prototypical band inversion from Sb \textit{s}- to Tl \textit{p}-character for the lower and \emph{vice versa} for the upper band.
When introducing proper element substitution generating \ce{SbPbO3},
the SOC-induced splitting and inversion is shifted right to the Fermi level due to
electron doping through Pb.
The topological nature of the systems is investigated through calculation of the associated invariants revealing that 2D \ce{SbPbO3} is a TI with $\mathbb{Z}_2 = 1$.
The dispersion of the robust, topologically protected edge states within the $\sim$\SI{0.2}{eV} SOC-induced gap is computed for both \ce{SbTlO3} and \ce{SbPbO3} in ribbons with zigzag (ZZ) and armchair (AC) edge terminations, revealing Dirac points located at the Brillouin zone boundary (ZZ) and at the zone center (AC).

This work provides a platform for investigating robust non-van der Waals 2D topological insulators and highlights promising prospects for topological junction states in appropriately designed non-van der Waals heterostructures.

\section*{Data Availability}
\noindent
The associated primary research data are made publicly available via the Rossendorf Data Repository (RODARE) at \verb|https://doi.org/10.14278/rodare.4591| ~\cite{Lokamani_Rodare_2026}.

\begin{acknowledgments}
The authors thank Johannes Wasmer and Tom Barnowsky for fruitful discussions.
M.L. acknowledges the Helmholtz Visiting Researcher Grant from HIDA for a research visit at PGI, FZJ.
R.~F. acknowledges funding by the Deutsche Forschungsgemeinschaft (DFG), project FR 4545/2-1, collaborative research center SFB 1415, Project C08, Project ID 417590517, and for the \enquote{Autonomous Materials Thermodynamics} (AutoMaT) project by Technische Universität Dresden and Helmholtz-Zentrum Dresden-Rossendorf within the DRESDEN-concept alliance.
S.B.\ acknowledges funding from DFG through CRC 1238 (project number 277146847, subproject C01).
G.M., D.W. and S.B. acknowledge funding from EuroHPC and the German ministry BMFTR via the MaX Centre of Excellence (grant agreement no. 101093374).
The authors thank the HZDR Computing Center, HLRS Stuttgart, the Paderborn Center for Parallel Computing PC$^2$, and TU Dresden ZIH for generous grants of CPU time.
\end{acknowledgments}

\appendix*
\onecolumngrid
\section{Layer group and Wyckoff positions of 2D candidates}

\begin{table*}[!htb]
    \begin{minipage}{.49\linewidth}
        \centering
        \begin{tabular}{cccc}
        \hline
        \multicolumn{4}{c}{2D \ce{AgBiO3}}  \\
        \hline
        \multicolumn{4}{c}{Layer group: p31m; $a=5.8016$~\AA} \\
        \hline
        Atom & Wyckoff & \multicolumn{2}{c}{coordinates} \\
        \hline
           Ag  &  a  &                & $z=\pm 3.0742$~\AA \\
           Bi  &  b  &                & $z=\phantom{\pm}0.000$~\AA \\
           O   &  c  &  $x =\pm 0.3820$ & $z=\pm 2.2513$~\AA \\
        \hline
        \end{tabular}
    \end{minipage}
    \begin{minipage}{.49\linewidth}
        \centering
        \begin{tabular}{cccc}
        \hline
        \multicolumn{4}{c}{2D \ce{NaBiO3}}  \\
        \hline
        \multicolumn{4}{c}{Layer group: p31m; $a=5.7249$~\AA} \\
        \hline
        Atom & Wyckoff & \multicolumn{2}{c}{coordinates} \\
        \hline
           Na  &  a  &                & $z=\pm 3.595$~\AA \\
           Bi  &  b  &                & $z=\phantom{\pm}0.000$~\AA \\
           O   &  c  &  $x =\pm 0.3788 $ & $z=\pm 2.2551$~\AA \\
        \hline
        \end{tabular}
    \end{minipage}
\end{table*}

\begin{table*}[!htb]
    \begin{minipage}{.49\linewidth}
        \centering
        \begin{tabular}{cccc}
        \hline
        \multicolumn{4}{c}{2D \ce{SbTlO3}}  \\
        \hline
        \multicolumn{4}{c}{Layer group: p31m; $a=5.405$~\AA} \\
        \hline
        Atom & Wyckoff & \multicolumn{2}{c}{coordinates} \\
        \hline
           Tl  &  a  &                & $z=\pm 2.655$~\AA \\
           Sb  &  b  &                & $z=\phantom{\pm}0.000$~\AA \\
           O   &  c  &  $x =\pm 0.3828$ & $z=\pm 1.130$~\AA \\
        \hline
        \end{tabular}
    \end{minipage}
    \begin{minipage}{.49\linewidth}
        \centering
        \begin{tabular}{cccc}
        \hline
        \multicolumn{4}{c}{2D \ce{SbPbO3}}  \\
        \hline
        \multicolumn{4}{c}{Layer group: p31m; $a=5.681$~\AA} \\
        \hline
        Atom & Wyckoff & \multicolumn{2}{c}{coordinates} \\
        \hline
           Pb  &  a  &                & $z=\pm 2.251$~\AA \\
           Sb  &  b  &                & $z=\phantom{\pm}0.000$~\AA \\
           O   &  c  &  $x =\pm 0.3620$ & $z=\pm 1.128$~\AA \\
        \hline
        \end{tabular}
    \end{minipage}
\end{table*}

\twocolumngrid

\end{document}